# Bayesian two-interval test

Nicolas MEYER[1,2], Erik-André SAULEAU[1,2]

nicolas.meyer@chru-strasbourg.fr, ea.sauleau@unistra.fr

[1]CHU de Strasbourg, GMRC Pôle de Santé Publique, F-67091 Strasbourg, France,

[2]Université de Strasbourg, CNRS, iCUBE UMR 7357, F-67000 Strasbourg, France,

**Corresponding author**

Nicolas MEYER,  CHU de Strasbourg, GMRC Pôle de Santé Publique, F-67091 Strasbourg, France, nicolas.meyer@chru-strasbourg.fr,

**Abstract**

**Background:** The null hypothesis test (NHT) is widely used for validating scientific hypotheses but is actually highly criticized. Assuming a prior distribution on a parameter of interest, Bayes' theorem allows the calculation of the *a posteriori* probability of the parameter's value, conditional on the data. Although Bayesian tests overcome several criticisms, some limits remain. We propose a Bayesian two-interval test (2IT) in which two hypotheses on an effect being present or absent are expressed as prespecified joint or disjoint intervals and their posterior probabilities are computed. The decision to accept one hypothesis is based on a probability threshold. The same formalism can be applied for superiority, non-inferiority, or equivalence tests.

**Methods:** The 2IT was studied for three real examples and three sets of simulations (comparison of a proportion and a mean to a reference and comparison of two proportions). Several scenarios were created (with different sample sizes), and simulations were conducted to compute the probabilities of the parameter of interest being in the interval corresponding to either hypothesis given the data generated under one of the hypotheses. Posterior estimates were obtained using conjugacy with a low-informative prior. Bias was also estimated.

**Results:** As the sample size increases, the probability of accepting a hypothesis when that hypothesis is true progressively increases, tending towards 1, while the probability of accepting the other hypothesis is always very low (less than 5%) and tends towards 0. The speed of convergence varies with the gap between the hypotheses and with their width. In the case of a mean, the bias is low and rapidly becomes negligible.

**Conclusions:** We propose a Bayesian test that follows a scientifically sound process, in which two interval hypotheses are explicitly used and tested. The proposed test has almost none of the limitations of the NHT and suggests new features, such as a rationale for serendipity or a justification for a "trend in data". The conceptual framework of the 2-IT also allows the calculation of a sample size and the use of sequential methods in numerous contexts.

**Trial registration**

Not applicable

**Background**

In the biomedical sciences, extensive use is made of statistical tests when attempting to validate scientific hypotheses about a pathological mechanism or a biological phenomenon using data collected during an experiment. Indeed, in the absence of access to the population of interest for technical, financial and/or ethical reasons, it is on the basis of such experimental data that the extrapolation to the population from which these data come, i.e., the inference, is made.

Since the work of the philosophers of the Vienna Circle [1], scientific knowledge has been constructed through deduction and not induction. It is considered that one cannot prove a hypothesis but only falsify it. Such falsification is achieved by comparing predicted data against the corresponding data observed during the experimentation carried out to test the hypothesis.

The most commonly used technique for making inferences is the null hypothesis test (NHT) [2]. The principle of the NHT consists of first posing a hypothesis called the "null" hypothesis (H0). In its most common form, this hypothesis states that the parameter of interest (a difference in means, a correlation coefficient, etc.) has some reference value, generally equal to 0. An experiment is then carried out, and a test statistic (for example, a Student *t*-statistic or a Pearson correlation coefficient) is calculated from the measurements made. This statistic is an *ad hoc* quantity whose distribution is known under the null hypothesis (Student's *t*-distribution in the two previous examples). The probability of obtaining a value equal to or greater than the observed value of the calculated statistic is then determined (the p-value). If this probability is lower than the commonly agreed-upon threshold of 5%, or sometimes less, it is considered that "the statistic is too high". Since the corresponding data have nevertheless been observed, it is deduced that the null hypothesis cannot be correct, and it is rejected in favour of an alternative hypothesis (H1) specifying an effect to be highlighted.

The NHT in all its forms has recently been the object of quite strong criticisms, both on its internal logic and, as a result, on its interpretation. These criticisms are not new, with the first of them dating from 1938 [3, 4]; however, beginning in the 1960s [5], then in 1980 [6, 7], and especially in the last ten years, these criticisms have become more insistent, notably following the crisis of reproducibility observed not only in the human sciences but also in the biological sciences [8, 9]. The limits of the NHT are indeed numerous.

In the biomedical field, the null hypothesis most often cannot be true *a priori*. It is inconceivable, either biologically or mathematically, that a difference in effect will be exactly equal to 0 (0.000...). More generally, it is impossible to say statistically that an effect is strictly equal to a given value (for example, a difference exactly equal to 2.5 units) since this amounts to testing whether a point has some finite thickness. Since a point-like null hypothesis can never be validated, it is meaningless to reject it.

The commonly used form of the NHT is in fact the result of the fusion of two incompatible theories, that of Fisher and that of Neyman and Pearson [10]. In this composite version of the test, the alternative hypothesis, which is essential to the theory of Neyman and Pearson, is often omitted, as proposed by Fisher, and the rejection of the null hypothesis leads to the simple acceptance of a difference between the groups, without an alternative hypothesis regarding this difference always being clearly specified beforehand. Rejection of the null hypothesis thus leads to acceptance of an undefined difference.

The NHT does not explicitly compare the two hypotheses, which would allow a true comparison of the relative merit of each hypothesis according to the observed data. Indeed, the test statistic is calculated only under H0 and not under H1. The probability of the data is not computed under H1. Therefore, H1 can only be accepted by rejecting H0 because the data are judged incompatible with H0, even though we know nothing about the probability of the data under H1.

The p-value is an index that depends on both the effect size and the sample size. Two identical p-values can therefore correspond to very different effect sizes, depending on the size of the samples, making it impossible not only to directly compare two p-values but also to judge the strength of an effect on the basis of the p-value alone. Based on this consideration, it becomes absurd to say that the lower the p-value is, the lower the probability that H0 is true [11]. The NHT and the confidence interval are both based on a point estimate of a parameter and an estimate of the variance of that parameter estimate. This variance is large when the sample size is small. Thus, when the sample size is small, the confidence interval for the parameter is wide, encompassing values that are not very credible. The width of the interval can nevertheless be compatible with a significant test; for example, consider an odds ratio (OR) of OR = 2.66 [1.19 - 5.97], p = 0.0175. However, in most realistic situations, values within this range that are greater than 3, let alone 4, would not be credible. Thus, a significant test on a small number of people suggests that the observed effect is large, but this observed value is then a poor estimate of the true value of the parameter because a significant test on a small number of people leads to overestimation of the real effect, with effects that are too large to be realistic [7]. Depending on the sample size, the test may therefore select effects of completely different magnitudes but from the same population. In other words, the current use of the NHT tends to define the size of the effect *a posteriori* in accordance with a probability whose value also depends on the sample size and not only on the size of the effect itself. The NHT thus allows one to declare the existence of effects that are not explicitly sought and, in fact, are mere overstatements of a true effect. To avoid this shortcoming, a statistical tool that could separate the influences of the sample size and the effect size would be extremely useful for testing hypotheses.

Another characteristic of the NHT is the arbitrary use of the same decision threshold (p = 0.05) in almost all situations. The use of an identical threshold in all cases can lead to contradictory conclusions when using a Neyman–Pearson test. A study with a p-value of 0.06 would conclude that there is no difference and would not reject the null, whereas another study testing the same data but with p = 0.04 would reject the null hypothesis, even though the confidence intervals would be very close. While this conclusion is understandable in the logic of the Neyman–Pearson test, it does not account for the great similarity between these two situations in terms of the effect size.

The combination of the two limitations mentioned above can lead to absurd conclusions. Let us imagine 3 studies evaluating the relationship between an exposure and a disease, with the following results: $OR_1$ = 2.66 [1.19; 5.97], p = 0.0175; $OR_2$ = 1.76 [1.00; 3.08], p = 0.049; and $OR_3$ = 1.62 [0.96; 2.83], p = 0.091. $OR_1$ and $OR_2$ would often be judged as leading to the same conclusion regarding the relationship between exposure and disease because their associated p-values are close, although their respective confidence intervals cover very different value regions. On the other hand, $OR_2$ and $OR_3$ would be considered to give opposite conclusions on the basis of the p-value, although their values are close and their respective confidence intervals greatly overlap [12]. Such conclusions are found very often in the literature, and it would therefore be valuable to have a tool that could avoid or at least limit the possibility of contradictory conclusions. Finally, it should be noted that the use of confidence intervals instead of p-values, although suggested by many authors, is not a satisfactory solution since the p-value and the width of the confidence interval are intimately linked [13].

By definition, the p-value is the probability of having data that are at least as far from what one would expect under H0 as the observed data are. This formulation implies that the NHT also considers data that were not, in fact, observed, i.e., all data whose deviations would be larger than the deviation observed in the study. When comparing two groups, one with a mean of 2 units and the other with a mean of 3 units, we are therefore taking into account not only a difference of 1 unit between groups but also differences greater than 1 unit (differences of 2 units, 3, 4, etc., up to infinity). It is not clear how something that has not been observed could be an argument for or against a given hypothesis.

These errors in the internal logic of the NHT make it almost useless from a practical point of view and at least prevent it from rendering the services expected by its users. Many other criticisms have also been levelled against the interpretation of NHT results. For example, it is always possible to design a sample of a sufficient size to reject a null hypothesis on the basis of some difference, however small (even if it is clinically irrelevant). The current, rogue use of the NHT is mainly to look for significant effects rather than necessarily for medically or biologically relevant effects. Another consequence is that the user of the test tends to equate a biological effect with a statistical effect, whereas a statistically significant effect may, in fact, be biologically insignificant and negligible, as revealed by a large sample size. Furthermore, when a result is "non-significant", it is often concluded that the sample size was insufficient and lacked power. This conclusion is a fallacy because it implies that one knew before the experiment that one could reject H0 and, therefore, that a difference truly existed between the populations on which the test was performed, even though the test was non-significant. If a difference is known to exist, there is no uncertainty about it, and therefore, there is no need to test it. In addition to this (especially as a consequence of the previous errors), there are many other possible errors of interpretation, more than 20 of which were listed in a paper hosted by the American Statistical Association in 2017 [11].

Although several authors have well illustrated the limitations of the NHT and its use, they have rarely proposed alternative solutions to all these difficulties. For example, a recent editorial in Nature [14] called for the elimination of p-values but remained rather vague on the procedures to be used to replace them.

It is often proposed that the p-value should be replaced by the confidence interval, but the p-value and the confidence interval cannot be dissociated from each other [13]. The use of confidence intervals to perform a test considering the magnitude of the effect has been proposed, corresponding to the notion of "magnitude-based inference", notably in psychology. However, this method was formulated in a frequentist framework, and the intervals are therefore intervals of predicted data, compatible with an effect, and not estimates of the probability that the effect of clinical interest is the correct one. The lack of application of the Bayesian methodology makes this approach obsolete [15].

Indeed, the NHT itself is based on the frequentist notion of probability, in which the probability of an event of interest is conceptualized as the frequency of observation of that event in a series of experiments during which the event is likely to occur. The corollary of this is that, in frequentist statistics, the parameters of interest (difference in means, etc.) have fixed, albeit unknown, values, whereas the data, via repeated sampling, can be used to asymptotically approximate the value of a parameter. This has a very important consequence: the NHT is, in fact, a prediction of data and therefore of the value of a parameter under $H_0$, followed by a comparison of the observed value with this predicted value. The reasoning then "works backwards" from this comparison to an inference on the hypothesis formulated with respect to the parameter of interest, which leads to confusion between the lack of specificity (the type I error risk, i.e., the probability that the test statistic lies beyond the threshold at the alpha level under the null hypothesis, since all calculations on the statistic are performed under this hypothesis) and the positive predictive value of the test (the probability that the alternative hypothesis is true given that the test is significant).

However, it is the positive predictive value that is of interest to the researcher: who would want to know what happens under the null hypothesis that a treatment has no effect? The real question is whether the treatment is effective conditional on the observed effect, and to answer this question, one must use the data to determine the (subjective) probability that the treatment is effective. However, to determine this positive predictive value, one must use Bayes' theorem based on a prior probability of each competing hypothesis being true before performing the experiment that will generate the data. It is easy to show that a significant test does not prejudge the truth of a hypothesis [11], and to overcome this drawback, it would be necessary to have a test procedure that is capable of introducing the *a priori* probability of each hypothesis and thus enabling the use of Bayes' theorem.

Indeed, the Bayesian construction of a test of a null hypothesis against an alternative hypothesis (the complement of the null hypothesis) follows easily from Bayes' theorem. If we can decide on an *a priori* probability of H0 (say p0), then the probability of H1 follows (1-p0). We observe data for which we can calculate two likelihoods, one under H0 and the other under H1. Bayes' theorem then allows the calculation of the *a posteriori* probabilities of H0 and H1 given the observed data. The literature then proposes several approaches for deciding which hypothesis to accept: essentially, the Bayes factor (BF), the deviance information criterion (DIC) and *a posteriori* probability calculation [16].

For example, one can construct a "maximum *a posteriori*" test [17], in which the hypothesis that has the highest *a posteriori* probability is accepted. A refinement is to construct a minimum cost test by assigning a cost to the decision error for each of the two hypotheses, similar to the α and β risks of frequentist tests [18].

Several authors have proposed the use of the BF as an alternative to the NHT [19], but, as has been well described by Kass and Raftery [20], the BF cannot in itself constitute a hypothesis test. The BF is the ratio of two likelihoods and is used to compare the relative (and only relative) merit of two hypotheses, regarded as two different models. However, regarding its use as a statistical test, it lacks any consideration of *a priori* probabilities, and even a large BF in favour of H1 over H0 does not necessarily imply a large posterior probability of H1.

Bayesian tests overcome several criticisms of the NHT, notably the difficulties of interpreting p-values and the invocation of unobserved data. Mengersen et al. [21] provided a fairly comprehensive list of measures for quantifying effect size. Furthermore, Makowski et al. [22] compared different indices of the existence of an effect in a Bayesian framework. However, all these measures are based on both the general principle of the NHT and a point-like null hypothesis. Moreover, it is currently a common Bayesian practice to calculate the probability that a parameter has a value above (or below) a certain threshold value (of clinical significance) from the *posterior* distribution of that parameter, which has the effect of mimicking the p-value of the NHT, including the adverse effect of selecting effect sizes that are too large. For example, in an epidemiological study, the credibility interval of the OR is typically calculated first, and then, the probability that the OR is greater than 1 (the reference value indicating the absence of a difference in effect between the compared groups) is calculated. Nevertheless, these tests are still based on the specification of a specific point-like null hypothesis and an alternative, complementary hypothesis. As we have already seen, in most situations, a point-like null hypothesis is not meaningful. Moreover, the calculation of the probability that the parameter has a value greater than a certain reference value leads to the selection of extreme values among those that reject the null hypothesis, which are not, in fact, very credible but nevertheless participate in the overall inference about the parameter. Finally, among the values of the parameter retained by the alternative hypothesis, some are, in fact, closer to the value of the null hypothesis than to a central value of the alternative hypothesis. This method therefore reproduces one of the limitations of the NHT. It can thus be concluded that what remains to be done to break free of the shortcomings of the NHT is to relax this constraint of a comparison to a point-like reference value.

To this end, Kruschke [23] proposed the "region of practical equivalence" (ROPE), effectively an interval null hypothesis, which he showed to exhibit good statistical properties. However, Kruschke's proposal neglected an important aspect of the test, namely, the formulation of the alternative hypothesis. All of the above considerations suggest that the alternative hypothesis should similarly be formulated as an interval specifying a "useful" effect with relevant values.

Within the classical framework, the probability of showing a non-null effect increases with the sample size, and the precision of the effect size also increases. Nevertheless, when the null hypothesis is rejected, the observed effect size is usually taken as the real value for the population, independent of the effect specified in the alternative hypothesis when parameterizing the NHT, although it is in fact an estimate of the former. Moreover, this effect size is considered the true value *after* the test, and thus, it cannot truly constitute a test since the assumption it is supposed to evaluate is defined afterwards. Although the precision of the effect size estimate increases with the sample size, the precision may still be insufficient. Finally, the probability that the effect lies in a given prespecified interval does not necessarily increase. Since the p-value has a unique value that depends on both the effect size and the sample size, it does not allow one to distinguish between the influences of these two factors, which significantly limits the interpretation of the results of a study. As Altman and Krzywinski [24] said, "any choice of results based on the outcome, rather than on prespecified hypotheses, will lead to selection bias".

Therefore, it would be useful to have a statistical tool that could unambiguously distinguish the role of the sample size from that of the effect size and test whether the effect lies in a prespecified interval, as the scientific method demands. The solution to this specific part of the problem is also to use a target effect interval.

We propose here a solution to the problem of hypothesis testing within a fully Bayesian framework and under a formulation that eliminates the shortcomings of the current methods, either the NHT and its derivatives or the current Bayesian formulation. The proposed solution consists of specifying two hypotheses expressed as intervals and computing the probability of each of these two hypotheses conditional on the observed data. The first hypothesis specifies the interval of values of the parameter of interest that would reflect a present and relevant effect, and the second specifies the interval of values that would characterize an absent or negligible effect. In the results section of this paper, the properties of this two-interval test (2IT) are presented using simulations. The application of the 2IT is illustrated with three examples. The advantages and disadvantages of this method are presented in the discussion.

**Methods**

In the planning stage of an experiment, scientific considerations make it possible, before any data are collected, to formulate a scientific hypothesis that we seek to validate. This scientific hypothesis implies the use of a parameter whose value, in the case of the 2IT, lies either in an interval that favours the scientific hypothesis or in an interval that disfavours it. These two intervals may be complementary or not. The effect sought is thus considered to be either present, thus confirming the scientific hypothesis, or absent, thus invalidating the hypothesis. These intervals are denoted by $H_P$ and $H_A$, respectively, representing ranges of values that support either the presence or absence of the effect sought. If θ is the parameter of interest (e.g., a difference in means), then the two statistical hypotheses (given the same names as the intervals) are:

$$H_P: \theta \in [\theta_{PL}; \theta_{PU}] \text{ and } H_A: \theta \in [\theta_{AL}; \theta_{AU}]$$

where L and U denote the lower and upper bounds, respectively, of each interval. We use $L_P$ and $L_A$ to denote the lengths of these two intervals.

It is important to note that a "present" effect does not imply a non-zero or high parameter value but means that the observed effect is in favour of the scientific hypothesis being tested. For instance, equivalence between two treatments will be considered present if the effect difference is near 0.

The decision to accept $H_P$ or $H_A$ is based on Bayesian techniques and the comparison of the *a posteriori* probabilities that the parameter value lies in each interval[1]. The principle of the 2IT implies the use of the subjective interpretation of probabilities to enable the comparison of intervals of parameter values conditional on the data. Thus, if the scientific hypothesis is true and supported by the data, the test will indicate an *a posteriori* probability of the parameter belonging to $H_P$ that is much higher than its probability of belonging to $H_A$. We set a probability threshold $\pi$ such that $H_A$ or $H_P$ will be accepted if $Pr(H_A|data) \geq \pi$ or if $Pr(H_P|data) \geq \pi$, respectively. Notably, $\pi$ should be set sufficiently high ($\pi > 0.5$) that it is not possible to accept both $H_A$ and $H_P$.

We use $Pr(H_A)$ and $Pr(H_P)$ to denote the prior probabilities of $H_A$ and $H_P$ and $Pr(H_A|D)$ and $Pr(H_P|D)$ to denote their posterior probabilities. A value of the ratio $Pr(H_P|D)/Pr(H_A|D)$ that is greater than 1 indicates that the data are in favour of $H_P$, and a value less than 1 indicates that the data are in favour of $H_A$. It is also possible to calculate $Pr(D|HP)/Pr(D|HA)$, the BF [20] for comparing $H_A$ and $H_P$.

The two intervals $H_A$ and $H_P$ can be joint or disjoint and can be of equal or different lengths, and $H_A$ can be (entirely) higher than $H_P$ or (entirely) lower. The limits of the two intervals are determined on the basis of either a theoretical model or bibliographic, physiological, physio-pathological or experimental knowledge, the latter of which is the most frequent case in the biomedical field.

With the 2IT, the same formalism can be applied to three classical test situations and the respective positions of the two intervals:

---

[1] The details of the calculations used in Bayesian techniques are not recalled here. See [25] or [17].

*Superiority* test: A superiority test is used when the scientific hypothesis being tested implies a greater value of the parameter of interest in one group than in another. For example, suppose that we wish to show that an experimental treatment E is more effective than a reference treatment R by testing whether the survival rate at 12 months with E is higher than that with R based on a hazard ratio HR. $H_P$ specifies the magnitude of the HR that indicates a result in favour of E's clinical superiority, and $H_P$ will be accepted if the *a posteriori* credibility interval of HR is included in $H_P$ with a probability greater than or equal to $\pi$: $\Pr(\theta \in H_P|D) \geq \pi$. The superiority of E will be rejected if $\Pr(\theta \in H_P|D) < \pi$ or if the posterior credibility interval of the HR lies entirely within $H_A$ with a probability greater than or equal to $\pi$: $\Pr(\theta \in H_A|D) \geq \pi$, where $H_A$ defines an HR value interval near 1.

*Non-inferiority* test: Suppose that we wish to show that an experimental treatment E is non-inferior to a reference treatment R. The $H_P$ interval specifies the set of values for which we will conclude biological or clinical non-inferiority, and we will conclude non-inferiority if $\Pr(\theta \in H_P|D) \geq \pi$. The $H_A$ interval specifies the set of values for which we will conclude that E is clinically or biologically effectively inferior to R, and $H_A$ will be accepted if $\Pr(\theta \in H_A|D) \geq \pi$.

*Equivalence* test: Suppose that we wish to show that an experimental treatment E and a reference treatment R led to biological results that can be considered equivalent. The $H_P$ interval specifies the magnitudes of the differences in parameter values that are in favour of the biological equivalence of E and R and therefore tend to support the equivalence hypothesis. Such an $H_P$ interval will reflect the presence of an equivalence effect by specifying values of E - R differences close to 0, within equivalence limits specified *a priori*. We will accept the equivalence hypothesis $H_P$ if $\Pr(\theta \in H_P|D) > \pi$. On the other hand, equivalence will not be confirmed and will, in fact, be invalidated if the value of the difference between the groups is within the $H_A$ interval, which specifies a difference between the two treatments that is too great for them to be considered equivalent. We will accept $H_A$, i.e., conclude that there is no equivalence, if $\Pr(\theta \in H_A | D) \geq \pi$.

In these three situations, we clearly are concerned with the presence or absence of an effect reflecting a scientific hypothesis and not with a null or non-null value of a parameter.

Specifying $H_P$ and $H_A$ involves specifying the bounds of the two corresponding intervals: $\theta_{PL}$ and $\theta_{PU}$ for $H_P$ and $\theta_{AU}$ and $\theta_{AL}$ for $H_A$. These values can be set independently of each other, but a few rules can simplify the process. For example, under the assumption that the values in $H_P$ are greater than the values in $H_A$, one can set the bounds $\theta_{PL}$ and $\theta_{AU}$ to the same value. Alternatively, we can set $\theta_{PL}$ and $\theta_{AU}$ to be halfway between the central value of $H_P$ and the central value of $H_A$. We can then fix $\theta_{PU}$ and $\theta_{AL}$ on the basis of symmetry about the central values of each interval. Of course, this way of proceeding is not necessarily the best in all situations, and it is not difficult to imagine intervals of different lengths ($L_P \neq L_A$) as well as asymmetric intervals around a target value, depending on the needs of the theory to be tested or the scale of the values (logarithmic, for example).

The lengths of the intervals should be determined by what can reasonably be expected to represent a result either disfavouring the scientific hypothesis (for $H_A$) or in favour of it (for $H_P$). The width of each interval ($L_P$ or $L_A$) quantifies the precision that is expected when confirming or refuting the scientific hypothesis. The credibility interval of the parameter of interest should therefore ideally be entirely contained within $H_P$ or $H_A$ to confirm or reject the scientific hypothesis being tested. The narrower the $H_P$ and $H_A$ intervals are, the more strict the test, but the more informative and decisive the conclusion on the scientific hypothesis.

The qualities of the 2IT were studied in three situations: estimation of a proportion, comparison of two proportions and comparison of a mean to a reference. In each situation, scenarios were created to reproduce the case in which $H_P$ was true or $H_A$ was true. Simulations were conducted to estimate the probabilities of concluding $H_P$ (the proportion of times within the set of simulations that the *a posteriori* credibility interval of the parameter was entirely included in the $H_P$ interval) when $H_P$ was true and when $H_A$ was true, and similar simulations were conducted for $H_A$. Thus, for each of the three situations and for each value of N, the empirical probabilities $Pr(H_P|H_P)$, $Pr(H_A|H_P)$, $Pr(H_P|H_A)$, and $Pr(H_A|H_A)$ were estimated. The *a posteriori* estimates were obtained in a conjugation situation (parameter of a binomial distribution or comparison of the parameters of two binomial distributions) or in a pseudo-conjugation situation (mean of a normal distribution).

*A proportion*

Several scenarios were simulated. The assumption $H_A$ of no effect was set to 4 different values of the reference proportion $\pi$, where $\pi \in \{0.1, 0.3, 0.4, 0.5\}$. The assumption $H_P$ of an effect was set to $\pi + \delta$, with $\delta \in \{0.1, 0.2, 0.3\}$. For each of the 12 scenarios, the sample size was increased from 10 to 1000 in steps of 10 (Figures 1-3).

In each scenario, the width of each interval was the same for both hypotheses and centred on the value of $\pi_X$. A binomial distribution $B(N, \pi_X)$, where X = A or P, was assumed for the data.

We computed the exact probabilities of $\pi$ lying in $H_A$ and $H_P$ conditional on $H_A$ and $H_P$ being true, i.e., $Pr(H_P|H_P)$, $Pr(H_A|H_P)$, $Pr(H_P|H_A)$, and $Pr(H_A|H_A)$, using a Jeffreys Beta(0.5, 0.5) prior distribution.

*Comparison of two proportions*

For the comparison of two proportions, $\pi_1$ and $\pi_2$, between two groups, two situations were studied: a superiority test and an equivalence test. In the superiority test, $\pi_1$ was set to 0.5, and $\pi_2$ was set to 0.5 (representing no difference, $H_A$) or 0.7 (representing a difference, $H_P$). In the equivalence test, $\pi_1$ was set to 0.5, and $\pi_2$ was set to 0.5 in the presence of equivalence ($H_P$) or to 0.7 in the absence of equivalence ($H_A$). The number N per group was varied from 20 to 800 for each group. In both cases, the $H_A$ and $H_P$ intervals were set to a width of 0.2 and centred on $\pi_1 - \pi_2$, i.e., corresponding to the interval [-0.1, 0.1] or [0.1, 0.3]. Due to the principle of the 2IT itself, the superiority and equivalence tests are in fact perfectly symmetrical; thus, the results are shown only for the superiority test, as the conclusion for the equivalence test is exactly the same, with the labels A and P permuted on each index.

A binomial likelihood was assumed for the distribution of a phenomenon in two populations, with true proportions P1 and P2. The *a priori* knowledge of these proportions was modelled according to Beta distributions with parameters of 0.5 and 0.5 (Jeffreys prior). The *posterior* distributions were therefore Beta distributions with parameters of $0.5+X_i$ and $0.5+Y_i$, where $X_i$ and $Y_i$ are the numbers of "successes" and "failures", respectively, in population $i \in \{1; 2\}$. $X_i$ and $Y_i$ were simulated using a binomial distribution $Bin(N, P_1)$ or $Bin(N, P_2)$. The *a posteriori* distribution of each proportion was estimated by assuming a Beta distribution for each proportion, using the conjugation property of Beta distributions with a Beta(0.5, 0.5) Jeffreys prior. The distribution of each *a posteriori* proportion was then obtained by drawing 100,000 random values from the *ad hoc* Beta *a posteriori* distribution, and a corresponding sample of 100,000 differences was then computed, of which the 2.5$^{th}$ and 97.5$^{th}$ percentiles were used to construct the bounds of the credibility interval of the difference of the two proportions.

We accepted the hypothesis $H_P$ ($H_A$) if the probability of theta lying in the $H_P$ ($H_A$) interval was greater than 0.95.

*Comparison of a mean to a reference*

A reference value of $\mu = 0$ was assumed. Data were simulated under the assumption that the effect is present if the mean is $\mu = 1$ ($H_P$) and the effect is absent if the mean is $\mu = 0$ ($H_A$).

The interval width was set to 1 unit, centred on 0 for H$_A$ or on 1 for H$_P$. For each situation, the number of simulations was 2000. The sample size N was increased in increments of 10 from 30 to 1000. In each iteration, the posterior mean and its credibility interval were estimated. H$_P$ or H$_A$ was accepted if the *a posteriori* credibility interval was entirely included in the H$_P$ interval (from 0.5 to 1.5) or in the H$_A$ interval (from -0.5 to 0.5), respectively. For each value of N, the empirical probabilities of accepting H$_P$ and H$_A$ conditional on H$_P$ and H$_A$ were calculated.

The *a posteriori* mean was calculated under the assumption of an unknown variance [25], with a normal distribution of mean µ$_0$ = 0 or 1 with the unknown variance set equal to 3 (with an *a priori* pseudo-sample size of 1). Random samples of size N were obtained from a normal distribution of mean 0 (under H$_A$) or 1 (under H$_P$).

The *a posteriori* mean µ$_n$ is then:

$$\mu_n = \frac{(\kappa_0/\sigma^2)\mu_0 + (n/\sigma^2)\bar{y}}{\kappa_0/\sigma^2 + n/\sigma^2}$$

and the posterior variance $\sigma^2$ is:

$$\sigma^2 = \frac{1}{\nu_n}\left[\nu_0\sigma_0^2 + (n-1)s^2 + \frac{\kappa_0 n}{\kappa_n}(\bar{y} - \mu_0)^2\right]$$

where $\bar{y}$ is the observed mean of the sample, $\kappa_0$ is the number of prior observations for the mean, $\nu_0$ is the number of prior observations for the variance, $\mu_0$ is the prior mean, *n* is the sample size and $\kappa_n = \kappa_0 + n$.

*Use of the 2IT in three situations from the literature*

The 2IT was also applied to three examples from the literature, the first involving the estimation of a relative risk (RR), the second involving the comparison of two means, and the third to demonstrate correct balance in a randomization process.

*Example 1*

In the work of Hajek et al. [26], a sample size was explicitly computed, making it easy to specify both assumptions in the 2IT. The goal of the study was to evaluate the 1-year efficacy of refillable e-cigarettes compared with nicotine replacement when provided to adults seeking help to quit smoking and combined with face-to-face behavioural support. The primary outcome was 1-year sustained abstinence. The sample size of 886 was calibrated to demonstrate an RR of 1.7 (23.8% abstinence rate in the e-cigarette group vs. 14% in the nicotine replacement group) with a power of 95%.

*Example 2*

McCann et al. [27] compared the neurodevelopmental outcomes at 5 years of age after general anaesthesia or awake-regional anaesthesia in infancy in an equivalence study. Their hypothesis was that there would be no clinically important differences in neurodevelopmental outcomes between general anaesthesia and regional anaesthesia. This was expressed as an expected range of equivalence of +- 5 points in the Wechsler Preschool and Primary Scale of Intelligence (WPPSI-III) Full Scale Intelligent Quotient (FSIQ) at 5 years.

*Example 3*

The 2IT can also be used to confirm a lack of difference when one might be observed. In a therapeutic trial, it is traditional to check that randomization has fulfilled its intended role by verifying the balance of the main characteristics of the subjects between groups. It is therefore expected that in 95% of cases, the difference between the groups will be non-significant. A non-significant test is most often interpreted as showing no difference, as a lack of evidence of difference [11], which is often misinterpreted as evidence of no difference. Zhang et al. [28] ran a trial comparing two chemotherapy regimens for recurrent or metastatic nasopharyngeal carcinoma in phase 3 trials. Table 1 in their paper compares the rates of subjects with recurrence with distant metastases at inclusion. The goal is to check that the randomization has worked properly.

**Results**

*Results of the simulations*

The performance of the 2IT was evaluated in three inference situations: estimation of one proportion, two proportions, and an average. Although these are three different situations, the overall findings are the same and are given in the next paragraph. The specific differences in each situation are described in a later paragraph.

For simplicity, we adopt $H_X$ and $H_Y$ to denote the hypotheses $H_A$ and $H_P$. The text should be read the same way for both cases. For example, in the case of one proportion, if $\pi_A = 0.5$ and $\pi_P = 0.7$, $Pr(H_P|H_P)$ is the same curve as $Pr(H_A|H_A)$ with $\pi_A = 0.7$ and $\pi_P = 0.5$, and *vice versa*. The curves in Figure 1b are therefore inverted with respect to the curves in Figure 1a.

*General conclusions, valid for all three situations*

The probability of accepting $H_x$ when $H_x$ is true, $Pr(H_x|H_x)$, progressively increases as the number of samples increases, tending towards 1. The probability of accepting $H_X$ when $H_Y$ is true, $Pr(H_X|H_Y)$, is always very low (less than 5%) and tends towards 0 as the sample size increases. The classification error rate is therefore very low and quickly becomes negligible. When the sample size is low, we most often conclude that there is a lack of power because the *a posteriori* credibility interval is wider than the target width, and the probability of accepting $H_x$ when $H_x$ is true is also very low. This effect confirms the ability of the 2IT to distinguish between a lack of power and a low sample size. The exact performance depends on the assumptions made about the values under $H_A$ and $H_P$. Thus, the speed with which $Pr(H_X|H_X)$ converges to 1 varies with the gap between $\theta_A$ and $\theta_P$ and with the widths of $H_A$ and $H_P$. However, the general pattern of the conclusions as described above is valid regardless of the situation. All of these conclusions are valid regardless of the widths of the $H_X$ and $H_Y$ hypothesis intervals and regardless of the gap between $\theta_X$ and $\theta_Y$. Only the rate of power growth and the gap between the levels of power for each assumption vary from one situation to another, while the general pattern is perfectly identical.

***Specific conclusions for each situation***

*For a proportion*

The results for one proportion (Figures 1-3) were obtained using the exact formulas. In these simulations, intervals of equal length were specified for $H_A$ and $H_P$. $H_X$ is accepted with a higher probability than $H_Y$ when $\pi_X$ is farther away from 0.5 than $\pi_Y$ is. This effect is expected in that the variance of a proportion depends on the value of that proportion, decreasing as the proportion moves away from 0.5. The power for $H_x$ will therefore be greater the farther the proportion specified in $H_x$ is from 0.5. Thus, in Figure 1, the curve representing $Pr(H_X|H_X)$ rises faster and earlier than the curve for $Pr(H_Y|H_Y)$ when $\pi_Y$ is farther from 0.5 than $\pi_X$ is. The effect is more important the larger the difference between $\pi_X$ and $\pi_Y$ is. When $\pi_A$ and $\pi_P$ are symmetrical with respect to 0.5 (for instance, 0.45 and 0.55), the probabilities of accepting $H_A$ and $H_P$ grow at the same rate, and the curves of $Pr(H_X|H_X)$ are perfectly superimposed.

Note that these configurations can correspond to both the case in which we are testing for equivalence ($H_A$: $|\pi_X - \pi_Y| \geq \delta$ and $H_P$: $\pi_X - \pi_Y < \delta$) and the case in which we are testing for a difference ($H_A$: $\pi_X = \pi_Y$ and $H_P$: $\pi_X \neq \pi_Y$).

*For two proportions*

The specific conclusions for the comparison of two proportions (Figure 4) are identical to those for the case of one proportion. The variance of the difference between two proportions depends on the value of each proportion, and the closer the proportions are to 0 or 1, the smaller the variance. The power for $H_X$ ($Pr(H_X|H_X)$) therefore increases faster than the power for $H_Y$ ($Pr(H_Y|H_Y)$) when $H_X$ specifies a difference that is farther from 0 than $H_Y$ does. The situation is reversed when $H_X$ specifies a difference closer to 0 than $H_Y$ does. When the absolute deviation from 0 is the same for $H_X$ and $H_Y$, the power curves match perfectly (these results are not very informative and thus are not shown).

For example, when $\pi_A = 0$ (zero difference between the two proportions) and $\pi_P = 0.2$, the power for $H_P$ is greater than the power for $H_A$.

*For a mean*

The general results described above are again found here. In particular, the power for $H_X$ increases as the sample size increases, and the risk of false positives or false negatives is virtually zero (Figure 5). In the case of a mean, the question of possible biases on the mean and variance arises. The simulations show that the bias on the mean is non-existent, with the sample mean being centred on $\mu_X$ regardless of whether $H_X$ is accepted. The details of this bias for N = 250 (figure 6) and N = 700 (figure 7) further show that the means of the samples retained under $H_X$ are centred on $\mu_X$ in a unimodal distribution. The non-retained samples are also centred on $\mu_X$ but in a bimodal distribution with, as expected, no means close to $\mu_X$; instead, the means of the samples are located towards the edges of the $H_X$ interval.

Concerning the variance, as with the NHT, one could expect that the $H_X|H_X$ conclusions would be partially linked to the selection of biased samples with lower-than-expected variances. Simulations have shown that this bias is present but of a magnitude similar to that found for the NHT and that the larger the sample size is, the smaller the bias. Table 1 to 4 shows the magnitude of the bias on the mean and the variance as a function of the sample size. The results that lead to the acceptance of $H_X$ under the assumption that $H_X$ is true have, on average, a smaller variance than expected. However, this bias is small and tends to be negligible in situation where the probability to accept $H_A$ or $H_P$ exceeds approximately 30%.

*Results for the three examples*

*Example 1*

For the example of Hajek et al. [26], using the 2IT formulation allows us to define two target intervals. In this context, the difference assumption ($H_P$) may be conveniently defined around the expected RR. Considering that the RR scale is not linear, the upper and lower $H_P$ target bounds are defined as $\exp(\log(1.7) / 2) = 1.30$ and $\exp(\log(1.7) + \log(1.7) / 2) = 2.22$, respectively. The $H_A$ interval is defined as [$\exp(-\log(1.7) / 2) = 0.77$; $\exp(\log(1.7) / 2) = 1.30$]. These intervals are asymmetrical around their expected RR values on the exponential scale but symmetrical on the log RR scale. The observed results reported by Hajek were 79/438 (18%) and 44/446 (9.9%) successes in the e-cigarette and nicotine replacement groups, respectively. The observed RR was 1.83 (1.30 - 2.58), p<0.001. Using the two interval hypotheses specified above with uniform priors, the probability that the observed RR is in H_P is 0.849, and the probability that the observed RR is in H_A is 0.028. The "missing" probability arises because the RR confidence interval is not centred on the target of 1.7, with the observed upper bound being larger than the H_P upper bound. The H_P probability shows that although the data are roughly in favour of the hypothesis of an effect, the assumption implied by the RR specified for the sample size computation is only weakly supported, with a probability of 0.849. Nevertheless, this hypothesis is 30.3 times more credible than the equivalence hypothesis. More data are thus necessary to ascertain that the real effect is an RR of 1.7 within the H_P margins, although the assumption of a larger success rate with the e-cigarette is credible. It must be stressed that the higher values in the confidence interval, above 2.22, are considered *too* large on prior grounds, despite showing a positive effect, and are thus less credible. The values above 2.22 indeed support an effect but do not support an RR of 1.7. One last point to be made regarding this example is that the probability ratio of 30.3 is larger than the 0.95-to-0.05 ratio that could be expected from a classical point of view, even though the probability of the most probable hypothesis is only .849. This suggests that H_A can be ruled out, while H_P still needs further data to be ascertained. Finally, there is some evidence of plausibility for a larger effect than the one defined in H_P, but to test this claim, a new hypothesis must be established and specifically investigated in a new study, perhaps using the actual H_P [1.30; 2.22] as an alternative hypothesis to be confirmed.

*Example 2*

In the 2IT formulation, the assumptions of McCann et al. [27] are expressed as H_P: -5 < mean FSIQ difference < 5 and H_A: |mean FSIQ difference| ≥ 5. The cited study observed a mean (sd) FSIQ of 99.08 (18.35) in one group and 98.97 (19.66) in the other group, with a mean difference of 0.23 [-2.59 - 3.06]. Using low-informative priors, the mean difference was evaluated as 0.10 [-3.39; 3.64], Pr(H_A) = 0.005, Pr(H_P) = 0.995, and Pr(H_P)/Pr(H_A) = 199;

thus, using a probability ratio threshold of 100, for instance, the prespecified equivalence can be considered present and is thus confirmed.

*Example 3*

In Zhang et al. [28], the rates of subjects with recurrence with distant metastases at inclusion were 131/181 (72%) and 119/181 (66%) in the gemcitabine group and in the fluorouracil group respectively. A non-significant Fisher test was interpreted as proof that the two groups were balanced on this criterion.

However, if we consider that a difference of proportion δ of at least 10% is reasonably suggestive of an imbalance, then the conclusion is different. In the 2IT formalism, the presence of imbalance is specified as $H_P$: |δ| > 10%, and the absence of imbalance is specified as $H_A$: |δ| < 10%. Using a uniform prior, the probability of $H_A$, i.e., that the between-group difference is within the [-10; +10] % interval, is only 0.764. Here, again, the non-significance of a test p-value was taken as a sign of no effect, and the effect size was not even specified, while there was in fact a probability of 0.237 of a proportion difference larger than 10%, suggesting the presence of imbalance.

In the same paper, the proportions of subjects rated with an Eastern Cooperative Oncology Group (ECOG) performance status of ECOG 0 in the two treatment groups were 59/181 (33%) and 62/181 (34%). The Fisher exact test p-value was 0.824. Considering here again that a difference of at least 10% between the groups is relevant, with the same $H_P$ and $H_A$ as before and using a uniform prior, the $H_A$ probability (that the between-group difference is within the [-10; +10%] interval) is 0.947, which is suggestive of real balance between the treatment groups in terms of this criterion.

***Consistency of results: return to the example of ORs***

In the introduction, we presented an example of three studies evaluating the relationship between an exposure and a disease in the form of an OR, with two examples of contradictory conclusions when comparing two of the three ORs. The use of the 2IT suppresses this type of contradiction.

Let us recall the results of these three studies:

- $OR_1$ = 2.66 [1.19; 5.97], p = 0.0175
- $OR_2$ = 1.76 [1.00; 3.08], p = 0.049
- $OR_3$ = 1.62 [0.96; 2.83], p = 0.091

Let us set $H_A$ = [0.9; 1.1] and $H_P$ = [1.1; 2.95], with a target OR of 1.8. Note that the $H_P$ interval is symmetrical on the log scale around log(1.8), which explains the asymmetry of this interval on the OR scale. Under a mild *a priori* assumption on the prior distribution of the OR (that it is located within the interval of [1/20 - 20]), the associated results of the same three studies are:

- $OR_1$ = 2.75 [1.16; 5.62], $Pr(H_P)$ = 0.018, $Pr(H_A)$ = 0.631
- $OR_2$ = 1.80 [1.00; 3.01], $Pr(H_P)$ = 0.053, $Pr(H_A)$ = 0.918
- $OR_3$ = 1.66 [0.92; 2.78], $Pr(H_P)$ = 0.090, $Pr(H_A)$ = 0.895

These results are perfectly consistent with each other and with the $H_P$ and $H_A$ hypotheses. The first two ORs give results that are weakly consistent with each other (unlike the NHT results) and provide different levels of support to $H_A$, whereas the last two ORs, which were considered discordant according to classical methods, are now highly consistent, with the $Pr(H_P)$ values being very close to each other and providing similar support to this hypothesis, with intervals that greatly overlap.

**Discussion**

We propose here a new way to perform a statistical hypothesis test. This test is formulated entirely in a Bayesian framework, and it explicitly considers two competing hypotheses, each expressed as an interval. Both hypotheses are explicitly used in the test procedure. At the end of the test, the probability of each hypothesis is obtained, which allows them to be compared unambiguously by quantifying the support for each hypothesis provided by the data, as allowed by Bayesian theory. The test thus produces a conclusion about each hypothesis. The proposed test has almost none of the limitations of the NHT and is therefore an interesting alternative to the NHT and a possible improvement to the inferential process in statistics. Moreover, it bypasses the limitations of other procedures formulated in either a classical frequentist or Bayesian framework that have been proposed as alternatives to the NHT.

Compared to the NHT, the 2IT restores an epistemological reality that is intellectually more satisfying because it allows one to affirm something in order to demonstrate its existence rather than affirming it by rejecting something else that seems not to be real. Moreover, as some scientific statements imply an absence of an effect, we provide here a way to formulate a solution to statistically support either the presence or the absence of an effect within a unified framework.

Regardless of the situation explored (one or two proportions or a single mean), our simulations show that the operational characteristics of the 2IT are satisfactory. The power of the test increases with the sample size, while the probability of falsely concluding the wrong hypothesis is small even for small sample sizes. The probability of accepting either $H_X$ ($H_A$ or $H_P$) when $H_X$ is true increases as the sample size increases, and the probability of accepting $H_X$ when $H_Y$ is true rapidly becomes very small.

The 2IT has a number of advantages. In contrast to the NHT, the 2IT does not guarantee that the hypothesis of an absence of an effect will necessarily be rejected if the sample is sufficiently large. One cannot sample to the point of "proving" one's favourite hypothesis, unlike in the NHT. One must instead reach sufficient precision for at least one of the two hypotheses or choose to stop including subjects, which is quite relevant and feasible in the context of Bayesian control of sequential analyses. When stopping before reaching the expected sample size and before either $H_A$ or $H_P$ is accepted, the conclusion will be that the sample is simply too small to warrant any conclusion. This makes it possible to distinguish a lack of power from an absence of an effect.

As it is based on Bayesian theory, the 2IT respects the principle of likelihood, which eliminates the paradoxes observed with the NHT, particularly in cases of multiple comparisons or sequential analyses [29]. Moreover, no extreme result is selected, and there is therefore no bias in the estimation. More precisely, the bias observed is of limited magnitude, decreasing as the sample size increases, and is of the same order of magnitude as that observed with conventional methods.

Various Bayesian or non-Bayesian methods using at least two intervals have also been proposed previously.

The method of Shih et al. [30], an extension of that of Goeman et al. [31], consists of a 5-region test for an OR. One of the regions corresponds to a negligible effect, and the other 4 regions correspond to moderate or large effects in favour of or against the exposure factor. However, the presented method is implemented entirely with a frequentist approach and therefore shares the numerous limitations of frequentist methods. Moreover, in the case of an OR, one of the intervals is an open interval extending up to infinity, which necessarily has the disadvantage of being able to select extreme effects, especially on small samples, most often falsely. Finally, its procedure is complex, with no fewer than 9 different possible conclusions. The interpretation is also complex and easily confused, especially since it only exacerbates the difficulties of interpretation of classical tests without providing a solution to the fundamental problem of hypothesis testing. Moreover, Ng [32] showed the limits of this method from a Bayesian point of view.

Campbell and Gustafson [33] proposed a conditional equivalence test, performed conditionally to the significance of a superiority test performed first. The two hypotheses to be tested therefore do not have the same status, and it is possible to have a non-significant test and then conclude both non-superiority and non-equivalence. This test shares all the defects of frequentist tests.

In the 2IT, as in the case of the ROPE [23], the $H_A$ hypothesis plays the role of the usual NHT null hypothesis, but it expresses the absence of the effect implied by the hypothesis in the form of an interval. This makes it possible to support this absence of effect since it does not take the form of a point-like value, which can never be exactly matched. However, the ROPE neglects the important aspect of the formulation of an alternative hypothesis. In the NHT, the alternative hypothesis is usually taken to be the complement of the null hypothesis. However, this often implies the consideration of implausible parameter values, such as an OR of 100 or 50. For small effects, as often observed in epidemiology, even an OR greater than 3 may be highly unlikely. This suggests that the alternative hypothesis should similarly be formulated as an $H_P$ interval specifying a relevant effect. A high probability of this $H_P$ hypothesis will tend to confirm it and thus support the presence of the effect sought.

The definitions of $H_P$ and $H_A$ and their associated lengths induce the possibility that the *a posteriori* credibility interval of the parameter of interest may be very precise, of a length less than both $L_A$ and $L_P$ but not entirely included in either $H_P$ or $H_A$ and therefore such that $Pr(\theta \subset H_X) < \pi$. This interval thus concentrates the values of the parameter in a zone of high probability at an unexpected location. In this way, the existence of a third zone is defined, which we will call $Z_S$. $Z_S$ is not associated with any scientific hypothesis since it has not been specified before the experiment. Its existence leaves room for an unexpected but nevertheless plausible and precise result, allowing us to evoke serendipity (hence the *S* of $Z_S$) to generate a new hypothesis that can then be formally tested via the 2IT in a new experiment. In this situation, in a manner somewhat similar to the classic use of the NHT, a hypothesis will be defined *a posteriori* from an interval of effect values with a high probability. Therefore, when faced with a conclusion of the $Z_S$ type, it is essential to validate the result by defining *a priori* a hypothesis to be validated in a new study and then calculating the probability of the effect reflected by this hypothesis, a hypothesis that will itself be confirmed or refuted in a "classic" 2IT. It is also assumed here that we are operating in a Bayesian framework in each stage of the process, with inclusion of the previous results in the *a priori* distribution for the second experiment. The hypothesis to be validated will express values close to those observed in $Z_S$, possibly modified or adapted to accommodate external knowledge that could refine the theory on $Z_S$.

The 2IT formulation finally gives meaning to the expression "a trend was observed in the data". Very often, when a result approaches significance but does not reach it, the authors speak of a "trend". This term is ambiguous in the context of the NHT because, as stated in the introduction, the NHT as it is commonly used is, in fact, a mixture of the Neyman–Pearson method, in which there is no place for the notion of a trend, and the Fisher method, in which a trend can be given a meaning. There are a great variety of formulations of this notion of a trend, but they all aim to express the idea that the hoped-for result has been approached but not yet reached; in general, such a conclusion is drawn in relation to p-values of approximately 0.10. We must remember here one of the major flaws of the NHT, namely, that it uses a p-value that dichotomizes the result into either the presence or absence of an effect, partly independently of the size of that effect, to define as present an effect size that has not been defined prior to experimentation and, therefore, without prejudging the size of this effect. There can thus be no trend in the result, or perhaps only a trend towards significance, which, as we have seen, should not be confused with the effect size. As the p-value does not quantify the effect size, the notion of a trend is not relevant.

On the other hand, the 2IT allows us to make sense of this notion. Indeed, as $H_P$ and $H_A$ specify an effect size, the closer the *a posteriori* probability of $H_P$ is to 1, the more likely the effect. When this probability is significant but not high, we can speak of a trend, since this trend will be defined by a level of certainty about the presence of an effect specified prior to the experiment. It is not a trend in the size of the effect, which is a target and therefore cannot change according to the data, but rather a trend in the level of certainty of an assertion about $H_P$. Insofar as the 2IT imposes a definition of the effect size sought before an experiment is carried out, when neither $H_P$ nor $H_A$ is supported and when the width of the interval of the parameter exceeds the target value, one can clearly say that the analysis lacks "power", if this notion can have any meaning in Bayesian methods. Here, again, insofar as the effect size is disconnected from the probability associated with it, the lack of an effect can clearly be linked to a too-small sample size.

To facilitate the analysis of test results in Bayesian inference, Mengersen et al. [21] has provided a list of indicators that quantify the probability of observing an effect of a certain magnitude. We cite here several measures of general interest: the estimate of the between-group difference, the credibility interval of this difference, the effect size measured by Cohen's *d*, the probability that Cohen's *d* exceeds a minimum value of clinical interest, and the predicted value for an individual in each group. However, here again, there is no explicit mention of two hypotheses specified before the study was carried out, much less of a comparison of two intervals, one specifying a negligible effect and the other a significant but reasonable effect, reflecting the expected effect. These measures are therefore extremely interesting, but they would be much more relevant when expressed within the 2IT framework and its definition of the hypotheses to be compared in the form of intervals.

With the 2IT, the question arises of how to choose the bounds of each interval. While the examples developed here illustrate two different ways of defining the bounds of the two hypotheses, others are also possible. However, whether the intervals are symmetrical or not, of identical or different lengths, joint or disjoint, defined by a physical theory or on the basis of expert opinion, the way in which these bounds are set is undoubtedly less important than the fact that these bounds must be set upstream of the experiment in order to support a confirmatory, and not merely exploratory, scientific scheme.

Once the experiment has been performed, the probabilities that the parameter of interest lies in the $H_P$ and $H_A$ intervals, i.e., $Pr(H_P|D)$ and $Pr(H_A|D)$, respectively, are calculated. These two probabilities can be compared in terms of their difference or, better, their ratio $Pr(H_P|D)/Pr(H_A|D)$. The higher this ratio is, the more strongly $H_P$ is supported. The closer this ratio is to 0, the more strongly $H_A$ is supported, and the less $H_P$ is supported, such that $H_P$ can be considered invalid. The comparison of this *a posteriori* ratio to the *a priori* ratio, $Pr(H_P)/Pr(H_A)$, allows us to estimate the BF, which quantifies the contribution of information from the data [20].

Whether we use the difference or the ratio, we are not proposing here to set a reference value as the decision threshold, as this would probably lead to the bad habit of always using the same threshold for all situations, which is obviously unwise. One cannot judge the promise of an anticancer drug using the same threshold as for evaluating the promise of a cold treatment or the role of a protein in the reproduction of fungi. That said, end users are often required to make decisions about whether to do something, such as a doctor who needs to decide whether to administer a treatment. Such binary decisions must be made on a case-by-case basis, by consensus within the learned societies. Any decision should therefore be made by the end user of the test statistic, not the statistician. Finally, a single scale for all situations would be absurd because the costs of the relevant decision errors differ greatly from one situation to another.

In the presentation of the method here, although it is explicitly and entirely based on Bayesian principles, we have not discussed in detail the role of the *a priori* distribution of the parameter of interest on the test conclusions. While the definition of an *a priori* distribution is essential to the formulation of the 2IT, which is based entirely on the Bayesian conception of statistical testing, the choice of this distribution has no influence on the underlying concept of the 2IT.

A "complete" Bayesian approach would also encompass the inclusion of a risk function in the 2IT. The use of such a risk function has not been developed here, but like the *a priori* distribution, this risk function does not modify and is not modified in any way by the 2IT principle.

The 2IT is able to distinguish between evidence of absence and absence of evidence, whereas the latter is often mistaken for the former with the NHT. With the 2IT, the absence of evidence is manifested by an excessively wide *a posteriori* credibility interval, while the evidence of absence is manifested by a confirmation of $H_A$. The 2IT also makes it possible to move beyond excessive concern with a significant test. This term, associated from the start by Neyman and Pearson and by Fisher with the demonstration of an expected effect, has perverted the use of statistical tests by giving the impression that only a significant test, rejecting H0, is interesting. This is particularly clear in the Neyman–Pearson framework. With the 2IT, either result, $H_A$ or $H_P$, can be affirmed at an agreed-upon risk, and thus, even a test concluding the absence of an effect can be classified as significant. Finally, a non-significant test will be simply and solely a test with insufficient power to affirm either the presence or absence of the effect sought. It will no longer be possible to confuse absence of evidence with evidence of absence, without needing to invoke significance. This method is therefore likely to greatly change the attitude of researchers regarding their publication strategies [34].

This paper presents the general principle of the proposed 2IT. Future work will focus on how to use this 2IT for sample size calculations, in the context of predictive probabilities, and in sequential analyses. Finally, the very concept of the 2IT makes it particularly useful in a meta-analysis.

Finally, the 2IT avoids the interpretative errors of the NHT and will probably be much easier to understand by non-statisticians, who are, after all, the final users of statistical tests. Because the 2IT directly concerns Bayesian inference and the specification of relevant intervals, clinicians will probably be more involved in the practical implementation of the test.

**Conclusions**

In the current context of intense discussions about the limitations of the null hypothesis test (NHT), many alternatives have been proposed, but these different proposals all have limitations and are quite impractical to use for some. It seems important to us to propose an alternative that could answer all, or at least a large part, of the criticisms that are levelled against these procedures. Our proposal seems to us to be the most complete one that is possible while providing a framework that is both simple and consistent with the needs of typical users of statistical tests. Recent discussions have even suggested a "ban" on p-values in science. This ban, even if it would lead to the hoped-for result, namely, the abandonment of a procedure that ultimately makes little scientific contribution, would nevertheless be difficult to implement in the absence of an efficient alternative. Bayesian methods, as they are often used at present, are also not entirely satisfactory because they do not completely follow

Bayesian concepts and instead sometimes attempt to mimic the principles of the NHT at the cost of preserving some of its shortcomings, in particular, threshold effects and the selection of extreme results based on an *a posteriori* observation of an effect not explicitly specified before the experiment. Moreover, it is not clear which authority would be able to enforce such a ban. On the other hand, it seems to us that our tool, because of its relevance and simplicity of use, would render the NHT *de facto* null and void, which would lead to its disappearance through practice rather than through a regulation that would not have much legitimacy in our scientific field in any case.

The proposed two-interval test (2IT) is thus a new alternative to the classical NHT in which the two hypotheses to be compared are explicitly stated and used in a consistent probabilistic framework, namely, Bayesian theory. Moreover, the 2IT provides a formal framework for serendipity. The conceptual framework of the 2IT also allows the calculation of a sample size and the use of sequential methods and can be adapted to numerous contexts. Although the use of this method remains to be described for more complex models (multivariate models, mixed models, and the Rasch model, for example) and particular situations (multiple comparisons, for example), as the proposed method is based on Bayesian theory, there are no conceptual difficulties but only, possibly, technical difficulties in the application of the 2IT to these models or situations.

The 2IT method is simple, straightforward, statistically sound and easy to implement. It is an alternative to both null hypothesis testing and tests of significance. It offers scientists "an alternative way of judging the reliability of their conclusions" [35].

**Declarations**

*Ethics approval and consent to participate*

**Not applicable**

*Consent for publication*

**Not applicable**

*Availability of data and materials*

**Not applicable**

*Competing interests*

**No competing interest declared.**

*Funding*

**None.**

*Authors' contributions*

**Both authors contributed equally to all elements of this work.**

**Not applicable**

# Appendix

The whole approach of the 2IT can easily be summarized as follows:

| **Insert in the text: the summary of the procedure:** |
|---|
| - Identify the endpoint, the associated random variable, the analysis model and the parameter of interest $\theta$ in the model |
| - Define the hypotheses of the presence and absence of the effect to be tested, $H_P$ and $H_A$, by indicating for each hypothesis the central value and the lower and upper bounds of the interval defining the hypothesis |
| - Define the *a priori* distribution of the $\theta$ parameter, fit the model, and estimate the point-like and posterior interval values of $\theta$ |
| - Compute the *a posteriori* probabilities that $\theta$ lies in the $H_P$ and $H_A$ intervals |
| - Compare the values of the two probabilities $Pr(H_P|D)$ and $Pr(H_A|D)$ by either their difference or their ratio |
| - Look for possible serendipity by comparing the length $L_S$ of the $\theta$ credibility interval with the lengths of $H_P$ and $H_A$ ($L_S < L_A$ or $L_S < L_P$) |
| - Conclude: $H_P$, $H_A$, serendipity, or insufficient power |
| - If serendipity is found, formulate a new hypothesis and plan an experiment to confirm it |

**References**

1. Popper KR. The logic of scientific discovery. London, New York: Routledge; 2002.

2. Neyman J, Pearson ES. On the problem of the most efficient tests of statistical hypotheses. Philos Trans R Soc Lond Contain Pap Math Phys Character. 1933;231:289-337.

3. Berkson J. Some difficulties of interpretation encountered in the application of the Chi-square test. J Am Stat Assoc. 1938;33:526-36.

4. Berkson J. Tests of significance considered as evidence. J Am Stat Assoc. 1942;37:325-35.

5. Rozeboom WW. The fallacy of the null-hypothesis significance test. Psychol Bull. 1960;57:416-28.

6. Cohen J. The earth is round (p<.05). Am Psychol. 1994;49:997-1003.

7. Freeman PR. The role of p-values in analysing trial results. Stat Med. 1993;12:1443-52; discussion 53-8.

8. Stodden V, Guo P, Ma Z. Toward reproducible computational research: an empirical analysis of data and code policy adoption by journals. PLoS One. 2013;8:e67111.

9. Lash TL. The harm done to reproducibility by the culture of null hypothesis significance testing. Am J Epidemiol. 2017;186:627-35.

10. Goodman SN. p values, hypothesis tests, and likelihood: implications for epidemiology of a neglected historical debate. Am J Epidemiol. 1993;137:485-96; discussion 97-501.

11. Greenland S, Senn SJ, Rothman KJ, Carlin JB, Poole C, Goodman SN, et al. Statistical tests, P values, confidence intervals, and power: a guide to misinterpretations. Eur J Epidemiol. 2016;31:337-50.

12. Gelman A, Stern H. The difference between "significant" and "not significant" is not itself statistically significant. Am Stat. 2006;60:328-31.

13. Feinstein AR. P-values and confidence intervals: two sides of the same unsatisfactory coin. J Clin Epidemiol. 1998;51:355-60.

14. Amrhein V, Greenland S, McShane B. Scientists rise up against statistical significance. Nature. 2019;567:305-7.

15. Sainani KL, Lohse KR, Jones PR, Vickers A. Magnitude-based Inference is not Bayesian and is not a valid method of inference. Scand J Med Sci Sports. 2019;29:1428-36.

16. Masson MEJ. A tutorial on a practical Bayesian alternative to null-hypothesis significance testing. Behav Res 2011;43:679-690.

17. Kruschke JK. Doing Bayesian data analysis: a tutorial with R, JAGS, and Stan. Boston, MA: Academic Press; 2015.

18. Hee SW, Hamborg T, Day S, Madan J, Miller F, Posch M, et al. Decision-theoretic designs for small trials and pilot studies: a review. Stat Methods Med Res. 2016;25:1022-38.

19. Tendeiro JN, Kiers HAL. A review of issues about null hypothesis Bayesian testing. Psychol Methods. 2019;24:774-95.

20. Kass RE, Raftery AE. Bayes factors. J Am Stat Assoc. 1995;90:773-95.

21. Mengersen KL, Drovandi CC, Robert CP, Pyne DB, Gore CJ. Bayesian estimation of small effects in exercise and sports science. PLoS One. 2016;11:e0147311.

22. Makowski D, Ben-Shachar MS, Chen SHA, Lüdecke D. Indices of effect existence and significance in the Bayesian framework. Front Psychol. 2019;10:2767.

23. Kruschke JK. Rejecting or accepting parameter values in bayesian estimation. Adv Methods Pract Psychol Sci. 2018;1:270-80.

24. Altman N, Krzywinski M. P values and the search for significance. Nat Methods. 2017;14:3-4.

25. Hoff PD. P values and the search for significance. London, New York: Springer; 2009.

26. Hajek P, Phillips-Waller A, Przulj D, Pesola F, Myers Smith K, Bisal N, et al. A randomized trial of E-cigarettes versus nicotine-replacement therapy. N Engl J Med. 2019;380:629-37.

27. McCann ME, de Graaff JC, Dorris L, Disma N, Withington D, Bell G, et al. Neurodevelopmental outcome at 5 years of age after general anaesthesia or awake-regional anaesthesia in infancy (GAS): an international, multicentre, randomised, controlled equivalence trial. Lancet. 2019;393:664-77.

28. Zhang L, Huang Y, Hong S, Yang Y, Yu G, Jia J, et al. Gemcitabine plus cisplatin versus fluorouracil plus cisplatin in recurrent or metastatic nasopharyngeal carcinoma: a multicentre, randomised, open-label, phase 3 trial. Lancet. 2016;388:1883-92.

29. Ryan EG, Brock K, Gates S, Slade D. Do we need to adjust for interim analyses in a Bayesian adaptive trial design? BMC Med Res Methodol. 2020;20:150.

30. Shih HY, Lee WC. A five-region hypothesis test for exposure-disease associations. Sci Rep. 2017;7:5131.

31. Goeman JJ, Solari A, Stijnen T. Three-sided hypothesis testing: simultaneous testing of superiority, equivalence and inferiority. Stat Med. 2010;29:2117-25.

32. Ng TH. Simultaneous testing of noninferiority and superiority increases the false discovery rate. J Biopharm Stat. 2007;17:259-64.

33. Campbell H, Gustafson P. Conditional equivalence testing: an alternative remedy for publication bias. PLoS One. 2018;13:e0195145.

34. Smaldino PE, McElreath R. The natural selection of bad science. R Soc Open Sci. 2016;3:160384.

35. Colquhoun D. The false positive risk: a proposal concerning what to do about p-values. Am Stat. 2019;73:192-201.

**Tables**

**Table 1**: Bias for the mean and the standard-deviation for N = 120.

**Table 2**: Bias for the mean and the standard-deviation for N = 150.

**Table 3**: Bias for the mean and the standard-deviation for N = 200.

**Table 4**: Bias for the mean and the standard-deviation for N = 400.

| N = 120 | n | mean | standard-error |
|---|---|---|---|
|  |  | [2.5% - 50% - 97.5%] | [2.5% - 50% - 97.5%] |
| $H_A$ | 17 | [9.96 - 10.01 - 10.02] | [2.51 - 2.66 - 2.77] |
| not $H_A$ | 1983 | [9.45 - 9.99 - 10.53] | [2.65 - 3.00 - 3.41] |
| $H_P$ | 20 | [10.97 - 11.00 - 11.03] | [2.47 - 2.64 - 2.74] |
| not $H_P$ | 1980 | [10.44 - 10.98 - 11.52 ] | [2.65 - 3.00 - 3.41] |

*Table 1*

| N = 150 | n | mean | standard-error |
|---|---|---|---|
|  |  | [2.5% - 50% - 97.5%] | [2.5% - 50% - 97.5%] |
| $H_A$ | 152 | [9.95 - 10.00 - 10.04] | [2.53 - 2.84 - 3.05] |
| not $H_A$ | 1848 | [9.54 - 10.00 - 10.48] | [2.66 - 3.00 - 3.33] |
| $H_P$ | 144 | [10.95 - 11.00 - 11.06] | [2.59 - 2.87 - 3.07] |
| not $H_P$ | 1866 | [10.52 - 11.00 - 11.49] | [2.66 - 3.01 - 3.37] |

*Table 2*

| N = 200 | n | mean | standard-error |
|---|---|---|---|
|  |  | [2.5% - 50% - 97.5%] | [2.5% - 50% - 97.5%] |
| $H_A$ | 592 | [9.90 - 10.00 - 10.09] | [2.68 - 2.97 - 3.23] |
| not $H_A$ | 1408 | [9.56 - 10.00 - 10.47] | [2.72 - 3.00 - 3.31] |
| $H_P$ | 636 | [10.90 - 11.00 - 11.09] | [2.67 - 2.96 - 3.24] |
| not $H_P$ | 1364 | [10.52 - 11.07 - 11.44] | [2.73 - 3.01 - 3.32] |

*Table 3*

| N = 400 | n | mean | standard-error |
|---|---|---|---|
|  |  | [2.5% - 50% - 97.5%] | [2.5% - 50% - 97.5%] |
| $H_A$ | 1678 | [9.81 - 10.00 - 10.19] | [2.77 - 2.99 - 3.20] |
| not $H_A$ | 322 | [9.62 - 10.21 - 10.38] | [2.80 - 3.01 - 3.21] |
| $H_P$ | 1651 | [10.81 - 11.00 - 11.19] | [2.78 - 2.99 - 3.20] |
| not $H_P$ | 349 | [10.61 - 11.20 - 11.40] | [2.79 - 3.02 - 3.20] |

*Table 4*

**Figure legends**

**Figure 1.** For a single proportion, the probabilities of accepting $H_A$ or $H_P$ when $H_A$ or $H_P$ is true, as functions of the sample size, for different scenarios in terms of the true values of $\theta_A$ and $\theta_P$. The intervals $H_A$ and $H_P$ are 0.1 units wide.

**Figure 2.** For a single proportion, the probabilities of accepting $H_A$ or $H_P$ when $H_A$ or $H_P$ is true, as functions of the sample size, for different scenarios in terms of the true values of $\theta_A$ and $\theta_P$. The intervals $H_A$ and $H_P$ are 0.2 units wide.

**Figure 3.** For a single proportion, the probabilities of accepting $H_A$ or $H_P$ when $H_A$ or $H_P$ is true, as functions of the sample size, for different scenarios in terms of the true values of $\theta_A$ and $\theta_P$. The intervals $H_A$ and $H_P$ are 0.3 units wide.

**Figure 4.** For the difference between two proportions, the probabilities of accepting $H_A$ or $H_P$ when $H_A: \delta = 0$ or $H_P: \delta = 0.2$ is true, as functions of the sample size. The intervals $H_A$ and $H_P$ are 0.1 units wide.

**Figure 5.** For a single mean, the probabilities of accepting $H_A$ or $H_P$ when $H_A: \theta_A = 0$ or $H_P: \theta_P = 1$ is true, as functions of the sample size. The intervals $H_A$ and $H_P$ are 0.5 units wide.

**Figure 6.** For a single mean, estimation of the bias on mean and variance for samples selected in $H_A$ or $H_P$ or not. The intervals $H_A$ and $H_P$ are 0.5 units wide. N = 250.

**Figure 7.** For a single mean, estimation of the bias on mean and variance for samples selected in $H_A$ or $H_P$ or not. The intervals $H_A$ and $H_P$ are 0.5 units wide. N = 750.

**Figures**

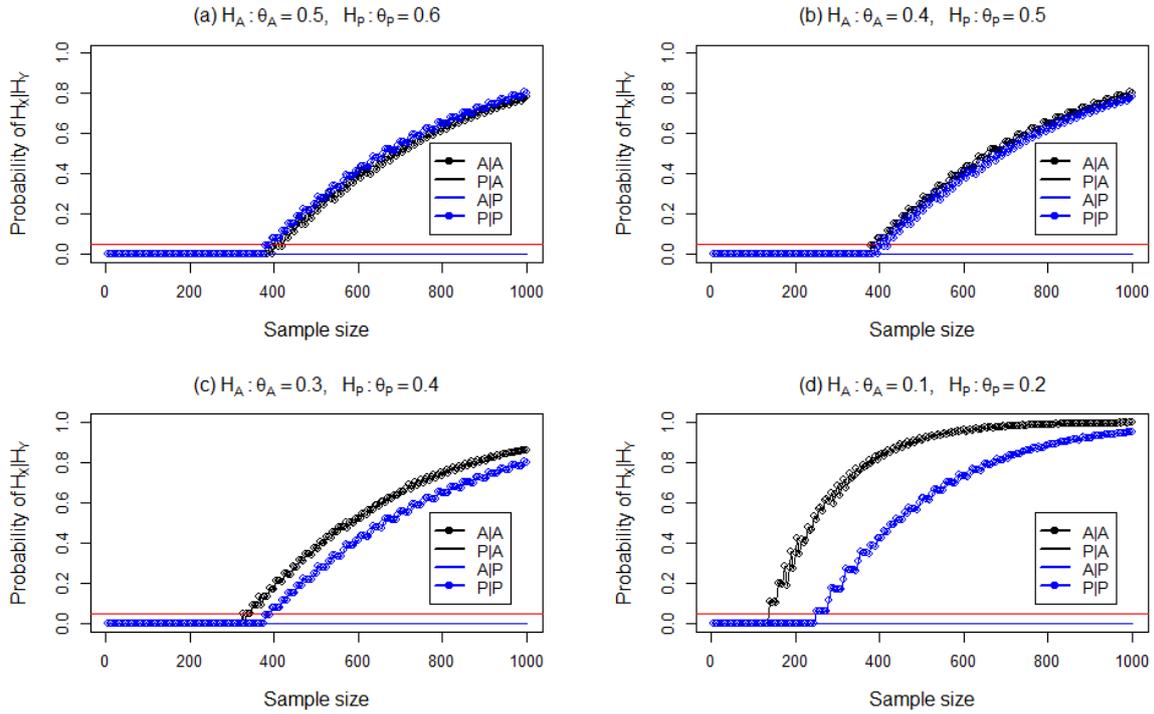

*Figure 1.*

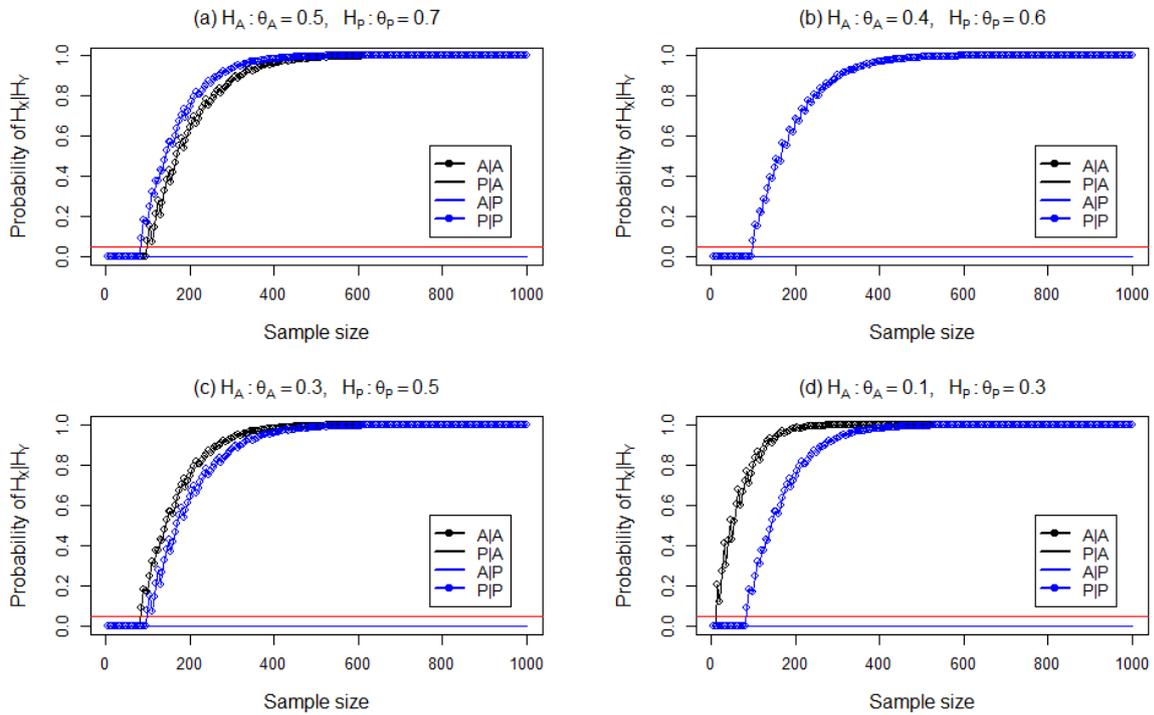

*Figure 2.*

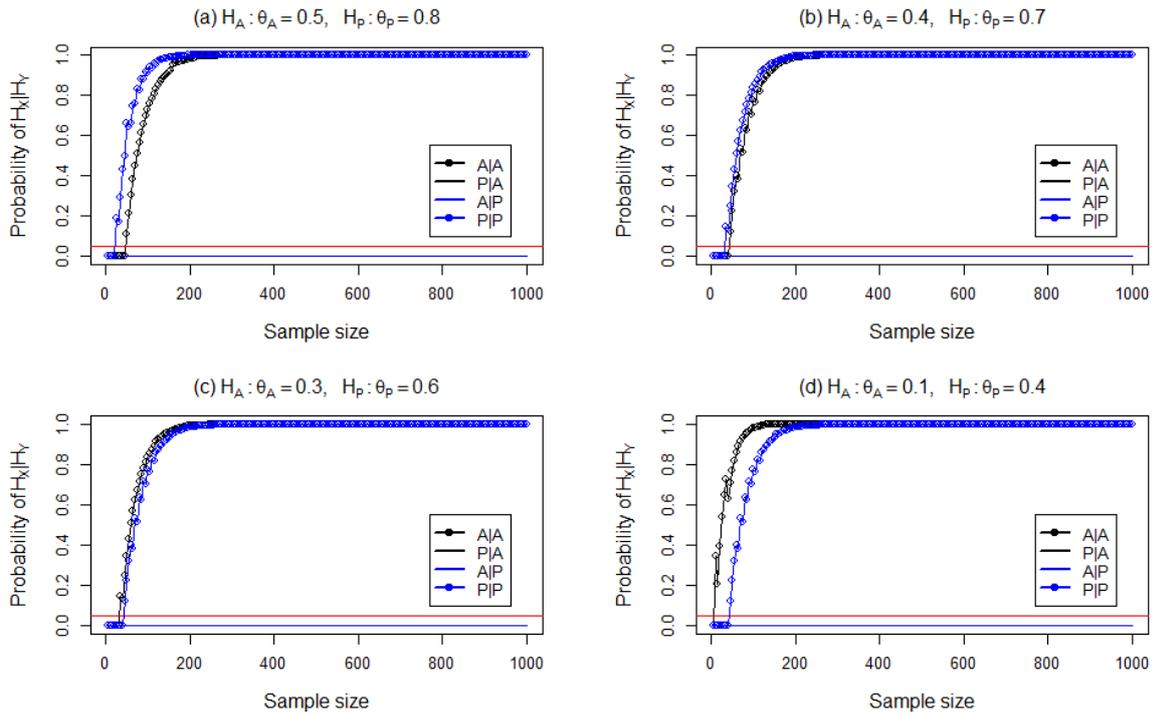

*Figure 3.*

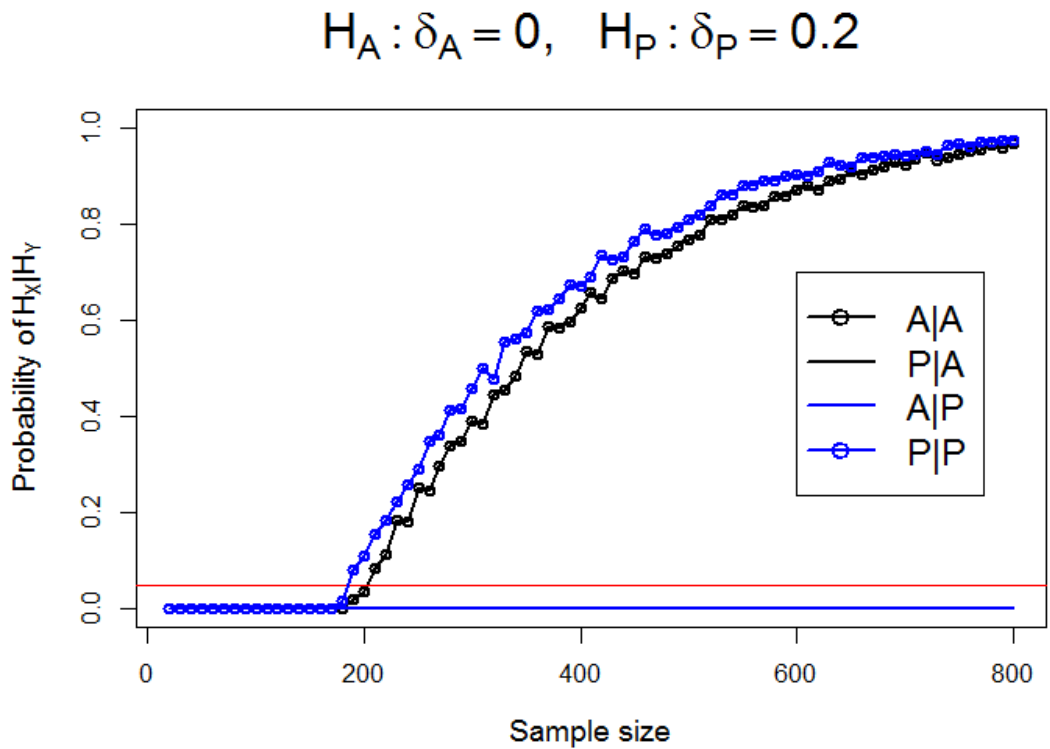

*Figure 4.*

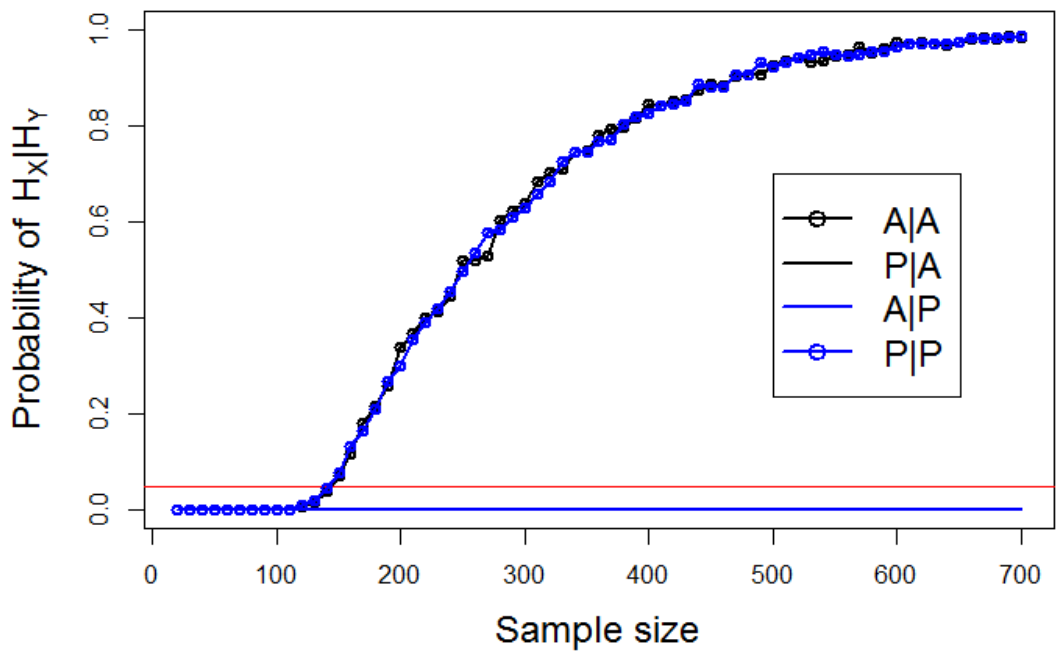

*Figure 5.*

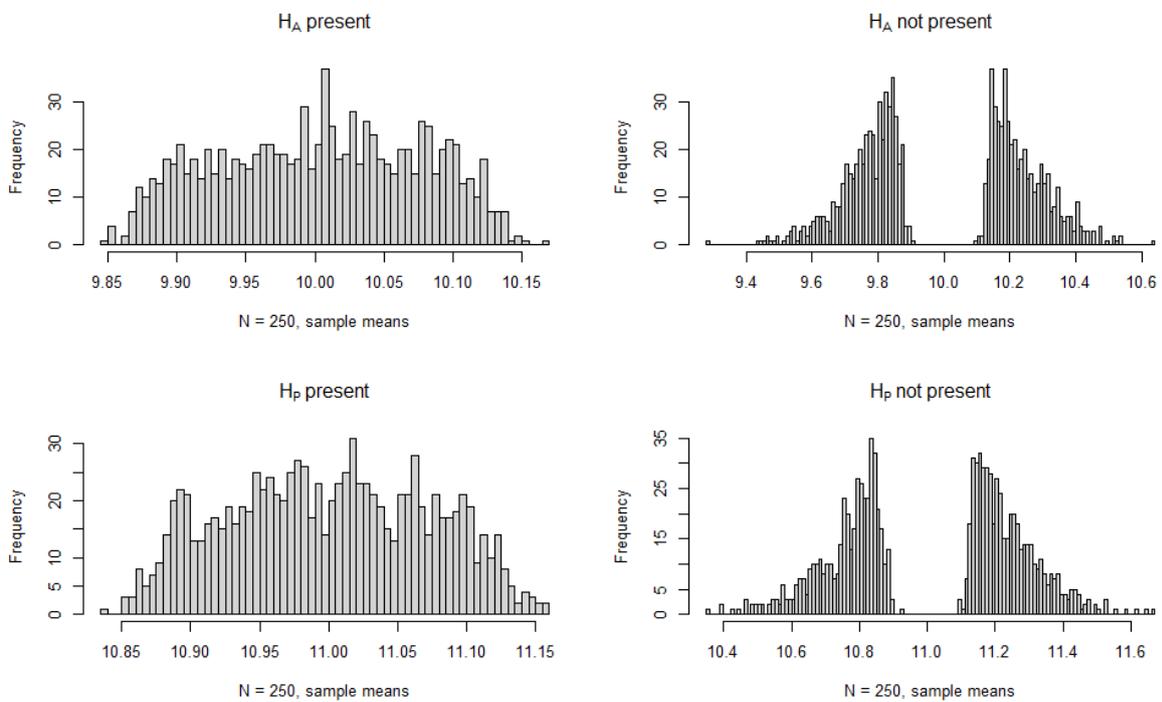

*Figure 6.*

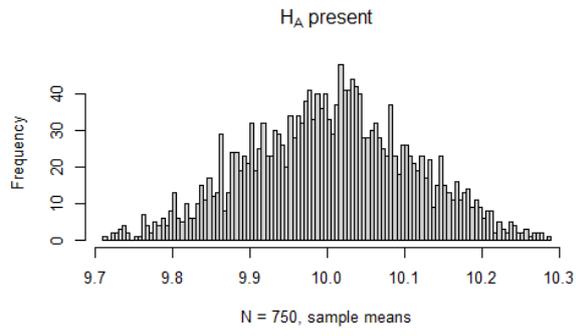 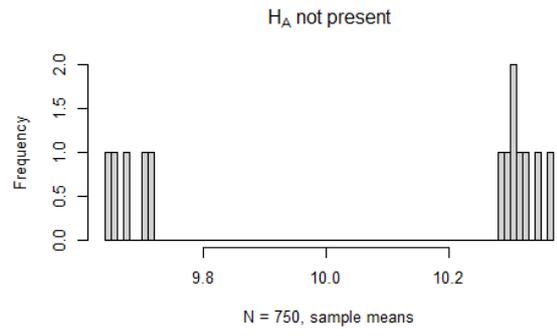
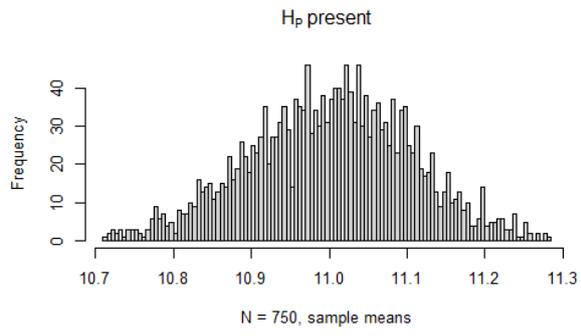 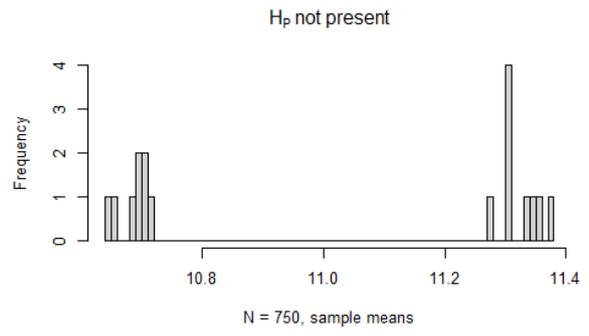

*Figure 7.*